Analysis of acoustic emission during the melting of embedded indium particles in an aluminum matrix: a study of plastic strain accommodation during phase transformation

Presented at "Atomistic Effects in Migrating Interphase Interfaces - Recent Progress and Future Study"

Michael M. Kuba, David C. Van Aken

Materials Science and Engineering
Missouri University of Science and Technology
223 McNutt Hall, 1400 N. Bishop, Rolla, MO 65409-0330, USA
Tel.: 573-341-4717
Email: dcva@mst.edu

**ABSTRACT**

Acoustic emission is used here to study melting and solidification of embedded indium particles in the size range of 0.2 to 3 μm in diameter and to show that dislocation generation occurs in the aluminum matrix to accommodate a 2.5% volume change. The volume averaged acoustic energy produced by indium particle melting is similar to that reported for bainite formation upon continuous cooling. A mechanism of prismatic loop generation is proposed to accommodate the volume change and an upper limit to the geometrically necessary increase in dislocation density is calculated as $4.1 \times 10^9$ cm$^{-2}$ for the Al-17In alloy. Thermomechanical processing is also used to change the size and distribution of the indium particles within the aluminum matrix. Dislocation generation with accompanied acoustic emission occurs when the melting indium particles are associated with grain boundaries or upon solidification where the solid-liquid interfaces act as free surfaces to facilitate dislocation generation. Acoustic emission is not observed for indium particles that require super heating and exhibit elevated melting temperatures. The acoustic emission work corroborates previously proposed relaxation mechanisms from prior internal friction studies and that the superheat observed for melting of these micron-sized particles is a result of matrix constraint.

## I. INTRODUCTION

Recent study of the aluminum-indium system has shown that equilibrium melting of the indium particles can be detected by acoustic emission (AE) techniques [1]. AE results from rapid energy release that creates elastic pressure waves in a material. According to literature, displacive solid-state transformations generate AE resulting from the shear mechanism of transformation and diffusive transformations normally occur too slowly to generate AE [2]. In steels, martensite [2] and bainite [3] generate AE, but formation of allotriomorphic ferrite or the eutectoid product pearlite does not [2]. Formation of Widmanstätten ferrite has been suggested to also generate AE [3]. Consequently, displacive or martensitic-like solid-state transformations are often distinguished from diffusion controlled phase transformations by the presence of AE [4]. However, liquid-solid transformations are also known to exhibit AE as the solid contracts, i.e. most materials exhibit AE upon solidification but not melting [5]. The exact cause of solidification AE is debated [6], but may be due to frictional noise between solid crystals [7], cluster addition or subtraction from the solid-liquid interface [8], or perhaps casting separation from the mold wall. AE is detected in crystallizing polymers due to cavitation in areas of occluded liquid where shrinkage stresses overwhelm the cohesive strength of the melt and void formation occurs [9].

Acoustic emission is also detected during tensile tests for dislocation creation and motion associated with a yield point drop [10] and for void nucleation at nonmetallic inclusions during ductile fracture processes [11]. However, even a small amount of prior cold work has been shown to drastically decrease the AE response from dislocation movement in aluminum during tensile tests [10]. Presence or absence of AE in aluminum is dependent upon the slip distance and a maximum pre-yield dislocation density of $2.34 \times 10^6$ cm$^{-2}$ for detectable AE upon yielding is predicted [10]. Thus presence of AE during phase transformations provides powerful insight into the mechanism of the transformation because the sources of AE are well documented.

Malhotra and Van Aken [12] have proposed a strain energy effect associated with the 2.5% volume change upon melting for embedded indium inclusions in aluminum. The calculated increase in melting temperature (~6 K) was in good agreement with superheat measured by differential scanning calorimetry (DSC) and that observed by internal friction. Malhotra and Van Aken also demonstrated that the internal friction peaks observed during melting are dependent upon applied test frequency and heating rate; and as a result, a matrix relaxation process that controls the degree of superheat was proposed [12]. Consequently, Kuba and Van Aken suggested that the stress conditions around an indium particle during melting are similar to that required for void nucleation and growth during ductile fracture [1]. The aluminum matrix was theorized to plastically accommodate the volume change of the melting indium particles by dislocation generation and motion. AE would be produced under these conditions when the matrix relaxes rapidly as calculated by Malhotra and Van Aken for indium particles sitting on grain boundaries [1]. The AE previously reported is duplicated in Figure 1 for reference. The AE detected is a function of indium content and the acoustic energy is plotted as the integral of the squared RMS voltage with respect to time.

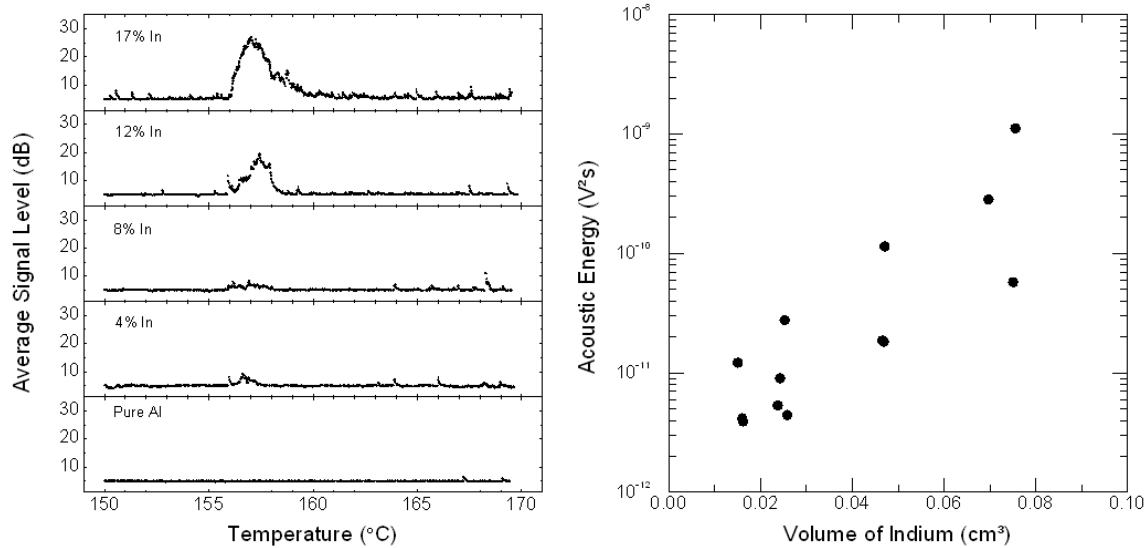

Figure 1. Acoustic emission and acoustic energy in as-cast aluminum-indium alloys as a function of composition. Left figure was originally published in Materials Letters [1].

Internal friction results from the work of Malhotra and Van Aken are shown in Figure 2 [12]. Two internal friction peaks were detected upon heating and three internal friction peaks were observed upon cooling. These melting and solidification events could be directly correlated with enthalpic changes observed using DSC. The two melting points detected were interpreted as a difference in diffusional relaxation times, but a dislocation generation model as suggested by Kuba and Van Aken to explain the AE was not considered. In contrast, the results of Wolfenden and Robinson [13] studying leaded brass produced only one internal friction peak at the melting temperature of lead. It should be noted that the strength of an internal friction peak is dependent upon the product of test frequency and relaxation time. A maximum in the internal friction peak is obtained when the product of the test frequency and the relaxation time is one. The lower test frequency used by Malhotra and Van Aken would be useful in probing diffusional relaxation mechanisms, while the 40 kHz test frequency used by Wolfenden and Robinson would be more likely to show the shorter relaxation time associated with dislocation generation mechanisms that might be detectable by AE. Interestingly, both internal friction studies were conducted for embedded inclusions that melted at homologous temperatures of 0.46 for the aluminum and 0.44 for the brass.

The initial discovery of AE in the Al-In system [1] was not predicted by literature and exposes a misunderstanding of phase transformations as they may relate to AE. The purpose of this study is to further investigate the nature of embedded particle melting via AE, since melting or solidification transformations produce a simple dilation that is common to most phase transformations, but is not complicated by long range diffusion or motion of a solid-solid interface. Previous studies have shown melting temperature to be dependent on particle location [12]; particles on grain boundaries melt at the equilibrium temperature, while particles embedded within aluminum grains melt at elevated temperatures. The current study investigates thermomechanically processed aluminum-indium alloys to show the effect of decreasing the number density of particles sitting on grain boundaries and observing the changes in AE. The study aims to further compile evidence of a dislocation-based relaxation mechanism of the volume strain associated with phase transformations.

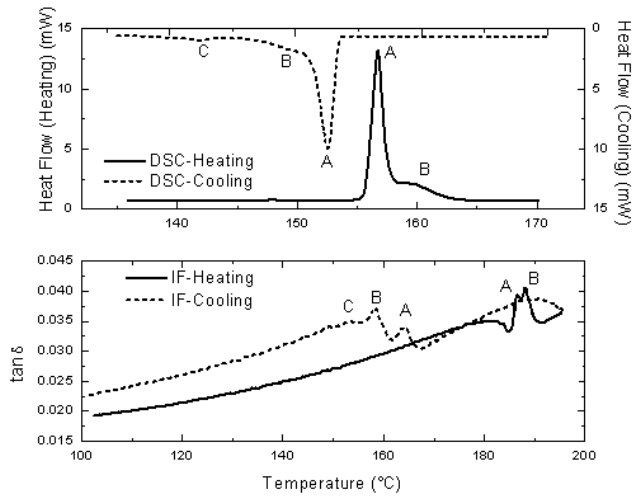

Figure 2. A comparison between internal friction (bottom) and DSC (top) of an Al-16In specimen.

## II. EXPERIMENTAL PROCEDURE
### A. Alloy Preparation

Aluminum specimens with nominal 17 wt.% indium additions were chosen for this study, since the composition is close to the monotectic composition and solidification produces a large number density of embedded indium particles. Compositions greater than 17.4 wt.% indium would result in liquid phase separation and a bimodal particle size distribution. The binary phase diagram is shown in Figure 3 for reference. Aluminum shot was melted in a fireclay crucible using a resistance furnace at 1093 K (820 °C). Aluminum-wrapped indium pieces were plunged into each melt and physically stirred for homogenization to encourage liquid miscibility and avoid liquid phase separation within the miscibility gap. Both materials were at least 99.99% pure with respect to metal content. The melts were chill cast into 13 mm and 14.3 mm diameter cylinders using aluminum molds. Two diameters were used so that specimens would be similar in size after the larger diameter specimen was mechanically swaged to decrease the diameter.

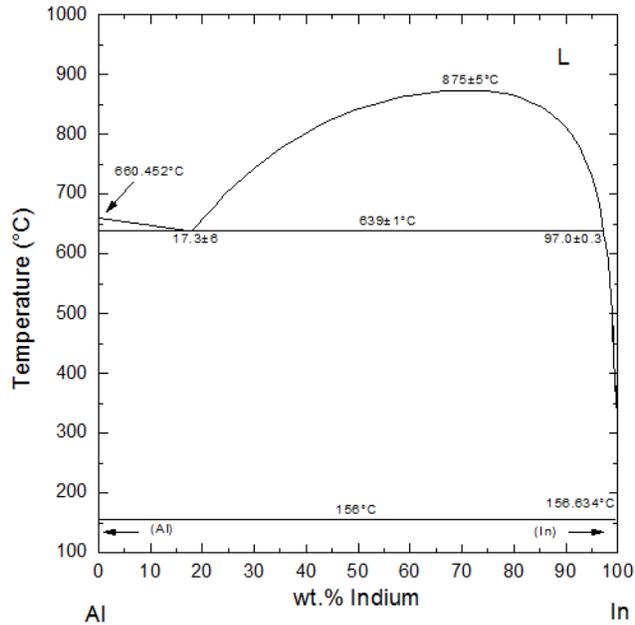

Figure 3. The aluminum-indium binary phase diagram showing a liquid miscibility gap and virtually no solid solubility for the solids. Originally published by Murray [14].

*B. Specimen Preparation*

The as-cast condition produced the highest indium particle concentration on grain boundaries. Cold work and low temperature annealing experiments were performed using material cast into the 14.3 mm diameter mold. Materials were tested in the as-cast and after annealing below the melting temperature of indium to ascertain the role of dislocation structures that may form during the initial indium particle solidification. Cast specimens and specimens swaged to 12.7 mm were annealed at 413 K (140 °C) for two hours. The annealing temperature was chosen such that the indium particles would not melt, but dislocation structures would recover during the heat treatment [15]. DSC and metallography were performed as described below.

Strain annealing was attempted to produce larger grain sizes and reduce the volume concentration of indium particles on grain boundaries. The 13 mm diameter cylinders were sectioned to 13 mm tall specimens and compressed to induce 5 to 6 percent plastic strain. The specimens were recrystallized at 773 K (500 °C) for one hour to produce larger grain sizes [16] where most of the indium particles will be embedded within the aluminum grains. Recrystallization was performed at 773 K (500 °C) and specimens were either air cooled or water quenched to determine the role of void formation upon rapid cooling. To study an intermediate condition, the 14.3 mm diameter cylinders were swaged to 12.7 mm to induce 20% strain. The cylinders were recrystallized at 523 K (250 °C) for 1 hour to produce a recrystallized grain structure between the as-cast and large-grained specimens [16].

*C. Characterization Techniques*

Specimens for optical microscopy were polished using standard metallographic procedures and electrochemically etched with Barker's reagent (1.8% $HBF_4$ in water) at 30 VDC. Five locations were used to determine grain size using Jeffries' planimetric method. The microstructure was further studied using a Hitachi S-570 scanning electron microscope. Five images per sample configuration were analyzed using ImageJ software to determine indium fraction and particle size. DSC was also used to characterize the melting and solidification phenomenon. Particle size was also investigated on the Al-17In alloy previously used for the AE study in reference [1]. A volume of dimensions 14 μm x 17 μm x 19 μm was examined by serial sectioning using a focused ion-beam SEM. Gallium ions were used to ion-mill the alloy. Micrographs taken during serial sectioning were aligned and recomposed into a three-dimensional volume using Avizo 7 software.

AE testing was performed on specimens machined to right cylinders with three orthogonal holes drilled through the specimen and normal to the surfaces. A resulting wall thickness of 4 mm was produced in order to minimize thermal gradients in the specimen. Temperature was recorded using a type K thermocouple swaged into a 2.25 x 2.25 mm hole with machining chips of the same composition for each specimen. The specimens were attached to 12.7 mm diameter aluminum alloy 6061 waveguides using a high temperature epoxy. For each test, a cross beam was mounted to the wave guide to suspend the specimen in a salt bath held at 473 K (200 °C). A PZT Navy type V transducer produced by Physical Acoustic Corporation was clamped to the end of the waveguide with Dow Corning high vacuum grease as a couplant. A virtual instrument designed in National Instruments LabVIEW software was used to record the AE signal and temperature simultaneously. After heating to the salt bath temperature, select specimens were removed from the salt bath and allowed to air cool while still monitoring AE and temperature. AE was measured as an average signal level with a time constant of 0.1 seconds as used by Van Bohemen [17]. Time-averaged AE data can be related to the energy of the transformation in a manner similar to heat flow using DSC, and measurement of AE is improved for continuous emission when the amplitude is low but the event occurrence rate is high. Specimens were subjected to a second heating and cooling cycle while AE was monitored, but with a 20 minute hold after heating to observe possible annealing effects. Thermal lag was removed from the AE plot by shifting the onset of AE in heating to the equilibrium melting temperature and upon cooling by direct comparison with DSC results. AE plots were exponentially smoothed to filter out noise.

Melt-spun aluminum with 12 wt.% indium alloy was also prepared to investigate the differences between indium nanoparticles and the micron-sized particles obtained during chill casting. DSC was performed to observe melting phenomena.

### III. RESULTS
*A. Particle Size Analysis*

Particle size analysis from the serial section and prior AE and DSC results from the same alloy [1] are shown in Figure 4 . A unimodal particle size distribution ranging from 0.2 μm to 3 µm in diameter was observed. In contrast, the DSC results show two distinct populations with different melting temperatures. Thus, it may be concluded that the melting behavior is not produced by two different size distributions. The size and shape of the indium particles was determined by examining the aluminum cavities that remained after liquation of the indium during ion milling. The gallium ions alloyed with and caused the indium particles to melt; the eutectic for binary indium and gallium is at 288.9 K (15.7 °C) and 21.4 wt.% indium. Consequently, minor pore broadening may have occurred. No particle faceting could be discerned from the cavity and internal voids within the indium would have been indistinguishable from those created by liquation. An image of the three dimensional reconstruction and an image from serial sectioning are shown in Figure 5.

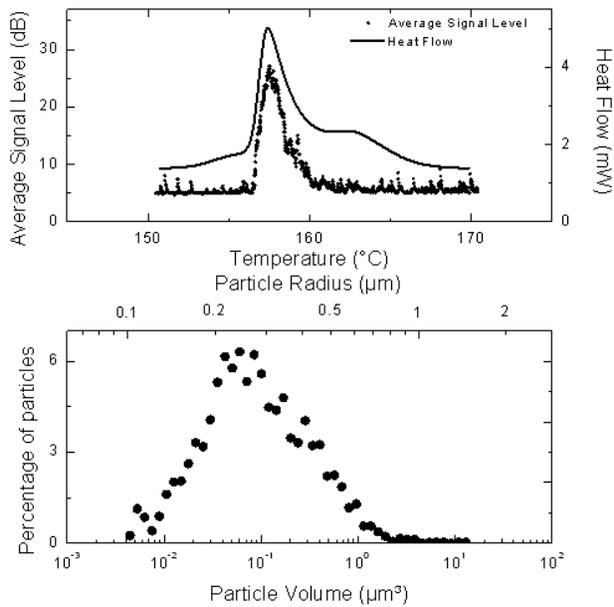

Figure 4. Particle count as a function of volume, as measured via serial sectioning. Particle size (bottom) is compared to prior results from Kuba and Van Aken [1] (top).

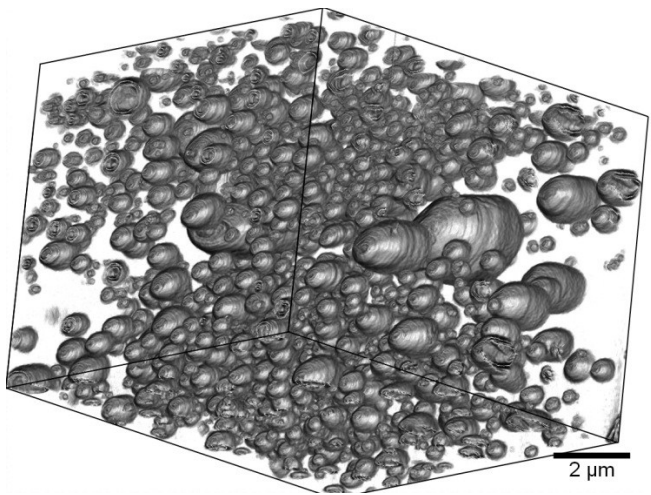
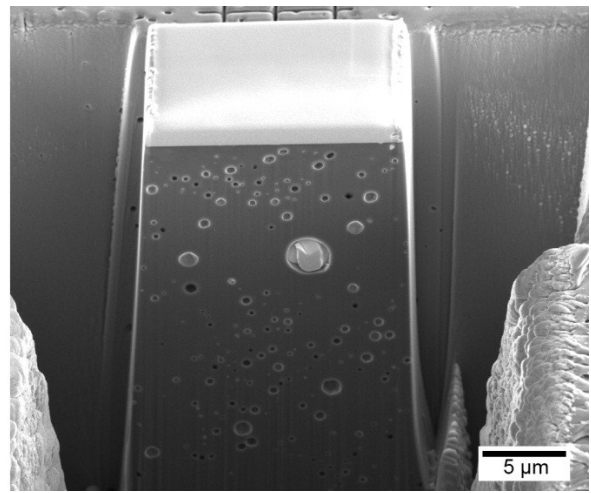

Figure 5. The three-dimensional reconstruction of the area selected for serial sectioning (left) and a selected image from serial sectioning (right).

*B. Quantitative Metallography*

Secondary electron images of polished specimens were used to measure the volume fraction of indium particles of the as-cast and thermo-mechanically processed materials. Results from quantitative metallography are presented in Table 1; uncertainties listed are at 95% confidence levels. The indium content was determined to be 15.6 wt.% based upon the measured volume fraction of indium particles and the reported densities of pure aluminum and indium. Representative optical and secondary electron micrographs are shown in Figure 6. Smaller grain size correlates with more grain surface area per unit volume, and it can be seen that the fraction of particles on the grain boundaries increases as grain size decreases. Grain surface area is emphasized here to quantify the fraction of indium on grain boundaries versus the fraction embedded within the aluminum grains.

Table 1. Quantitative Metallographic measurements of microstructural features of the Al-In alloys.

|  | Average Particle Diameter (μm) | Aluminum Grain Surface Area per Unit Volume (mm$^{-1}$) | Area Fraction of Particles at Grain Boundaries |
|---|---|---|---|
| As-cast | 1.16 ± 0.06 | 10.7 ± 1.6 | 0.33 |
| Intermediate-grained | 1.00 ± 0.02 | 11.5 ± 0.9 | 0.25 |
| Large-grained | 1.38 ± 0.04 | 7.54 ± 0.72 | 0.13 |

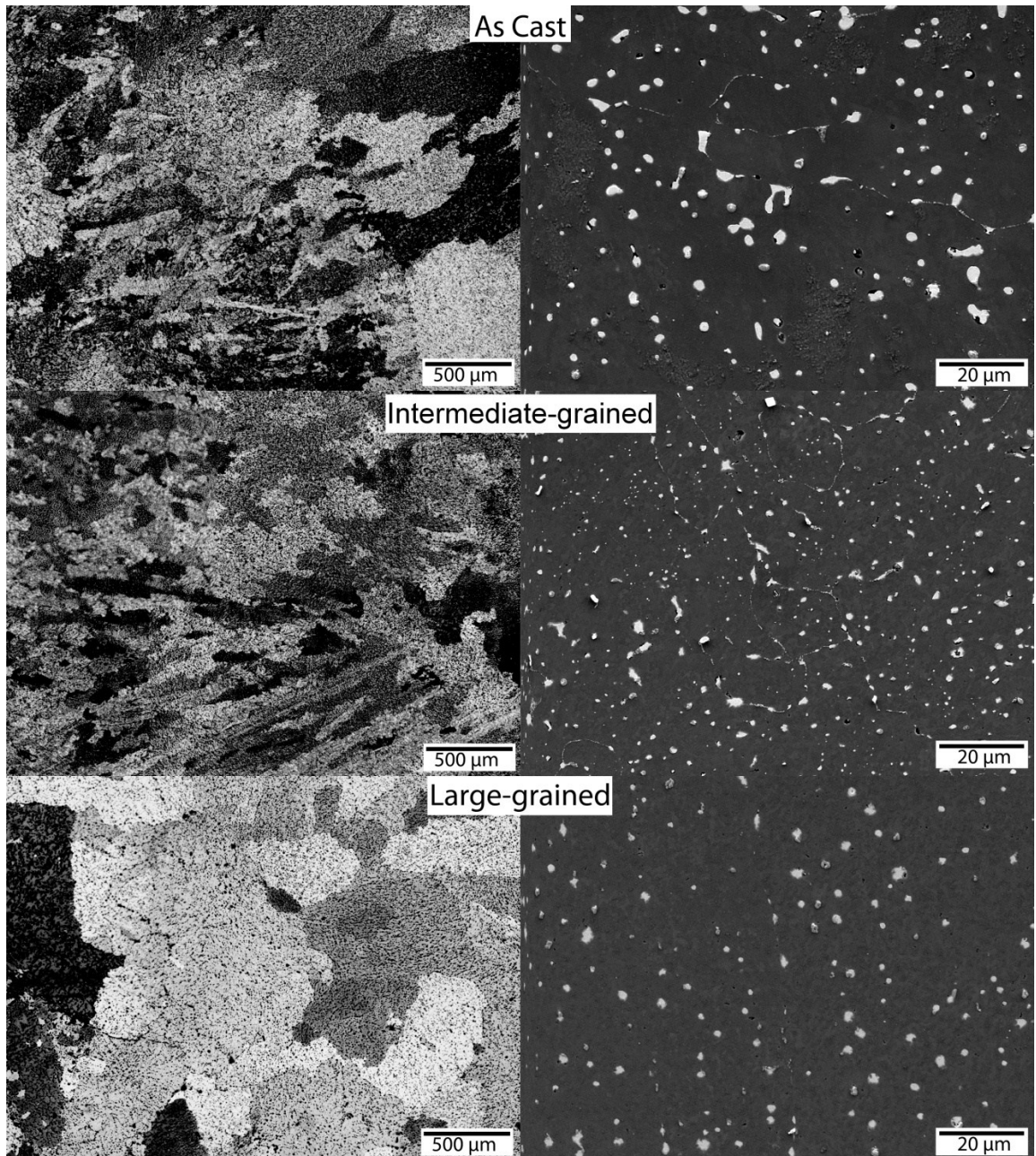

Figure 6. Optical (etched with Barker's reagent) and secondary electron images of the Al-In conditions studied. Backscattered electrons interacting with the pole piece result in atomic weight contrast in the secondary electron images.

*C. Acoustic and Calorimetric Studies*

The as-cast AE response is shown in Figure 7. Two peaks are present in the DSC in heating; one near the equilibrium melting temperature of 429 K (156 °C) [18], while a broader elevated-temperature peak partially overlaps the first. The as-cast condition demonstrates more superheat and less AE than previous results referenced in Figure 4. This may be related to the difference in average particle size: 1.16 μm presently compared to 0.68 μm previously. Low-temperature annealing experiments were performed to investigate matrix recovery (elimination of point defects and dislocation recovery) in determining the DSC and AE characteristics. AE and DSC comparisons for the low-temperature annealing test specimens are also shown in Figure 7 and are compared to the as-cast specimen. The 413 K (140 °C) anneal did not significantly change the DSC response in heating, but did increase the AE peak. Cold working the specimen followed by the annealing treatment is seen to decrease the AE relative to the as cast condition, but increases relative to the height of its own equilibrium melting DSC peak.

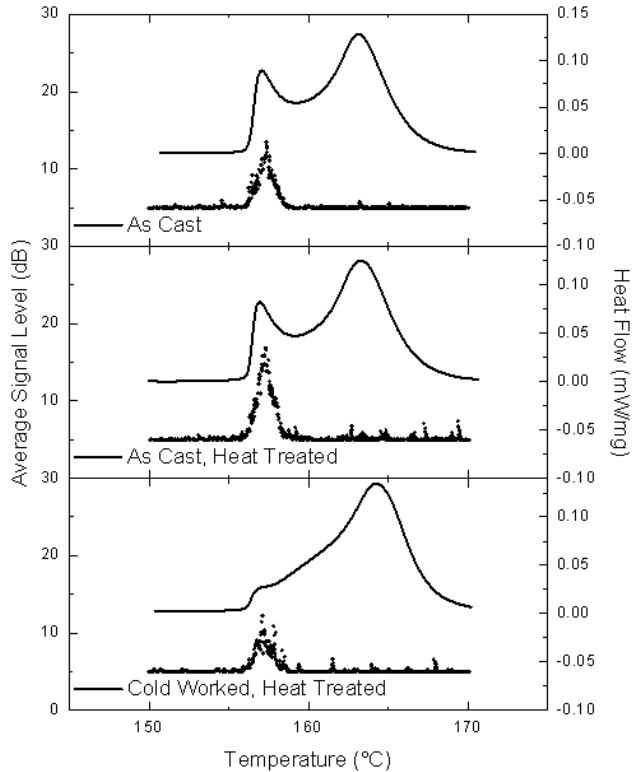

Figure 7. A comparison of DSC (line) and AE (scatter) in heating for the low-temperature anneal specimens, which were annealed at 413 K (140 °C) for two hours.

Two methods of deformation and recrystallization were employed to alter the grain size and thus effect the distribution of the indium particles in the test coupons. The largest aluminum grain structures were obtained by a technique of strain annealing where a small critical strain, sufficient to induce recrystallization, was employed with a high annealing temperature (500 °C) to produce a large grain structure. An intermediate grain structure was produced by 20% deformation and annealing at 250 °C. Characteristic AE and DSC comparisons for heating the various grain structures are shown in Figures 8 through 10. Figure 8 shows the AE and DSC results for the intermediate grain size. As shown in Table 1 the fraction of indium particles at grain boundaries decreased from 0.33 to 0.25. The proportion of indium melting at elevated temperature increased with increasing grain size and follows the trend reported by Malhotra and Van Aken. The proportion of indium melting upon the second heating cycle produced both an increase in AE and in the amount of indium melting near equilibrium. The DSC of the second heating also shows a melting point depression for a portion of the indium. The most dramatic shift in melting distribution was obtained for the strain annealed material having the largest aluminum grain size where the fraction of indium particles sitting on aluminum grain boundaries is only 0.13 (see Figure 9). Again upon heating a second time the amount of indium melting near equilibrium increased and again a melting point depression is observed. However, AE is only observed for those indium particles melting near equilibrium. Surprisingly, the AE and DSC results for the strain annealed material exhibited a strong dependence upon cooling rate from the 500°C annealing temperature. Specimens water quenched from the annealing temperature are shown in Figure 10. Here, the equilibrium melting peak height increased in proportion to the grain size whereas the measured AE peak was diminished and is absent for the second heating cycle. We suspect void formation during quenching and provide a complete analysis in the paper discussion.

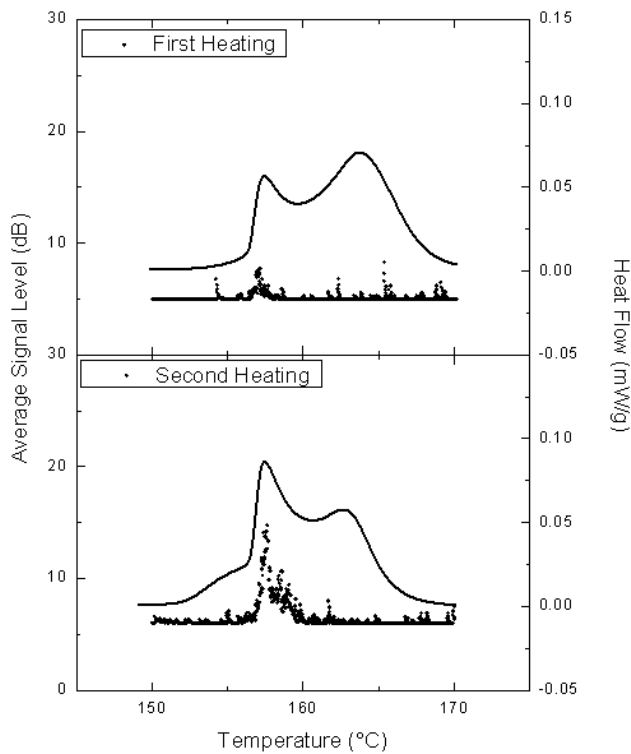

Figure 8. A comparison of DSC (line) and AE (scatter) for the first and second heating cycles of the intermediate-grained specimen, which was swaged 20% and recrystallized at 523 K (250 °C) for one hour.

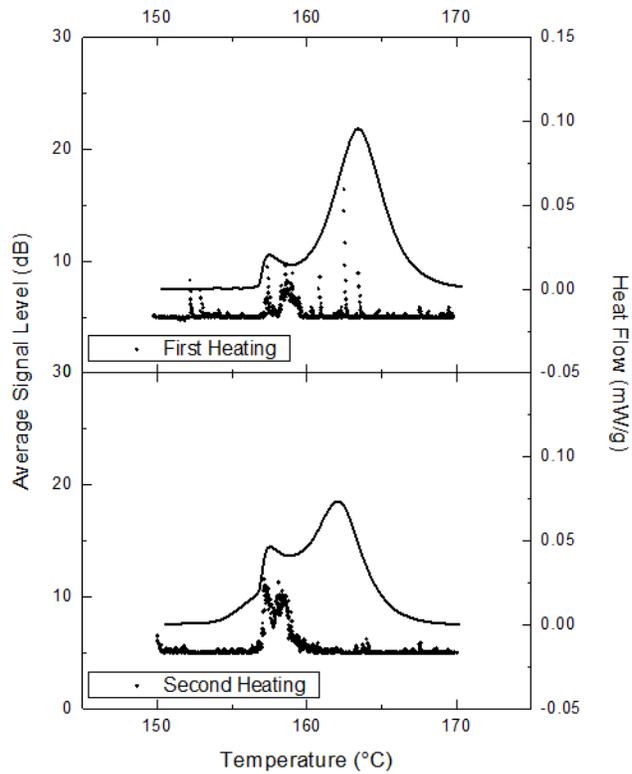

Figure 9. A comparison of DSC (line) and AE (scatter) for the first and second heating cycles of the air-cooled large-grained specimen, which was compressed 5% and recrystallized at 773 K (500 °C) for one hour.

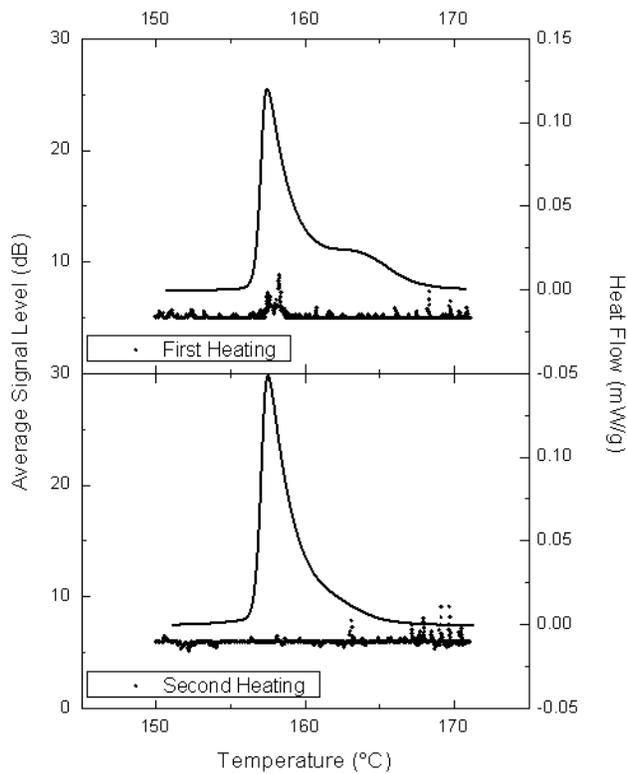

Figure 10. A comparison of DSC (line) and AE (scatter) for the first and second heating cycles of the water-quenched large-grained specimen, which was compressed 5% and recrystallized at 773 K (500 °C) for one hour.

The water-quenched large-grained specimens were also monitored in cooling for both AE and DSC, as seen in Figure 11. Several peaks were present in both the AE and DSC in cooling. All heat flows are graphed as being positive for the purpose of presentation and to aid the reader in comparing AE with DSC results. Good AE-DSC peak temperature correlation was obtained in cooling, with the two lowest temperature peaks coalescing into one single low-temperature solidification peak after the 20 minute hold at 473 K (200 °C). The relationship between the height of the AE and DSC peaks in cooling is not constant. While the DSC peaks are relatively similar in height on the first cycle, the AE is strongest for the two most undercooled peaks.

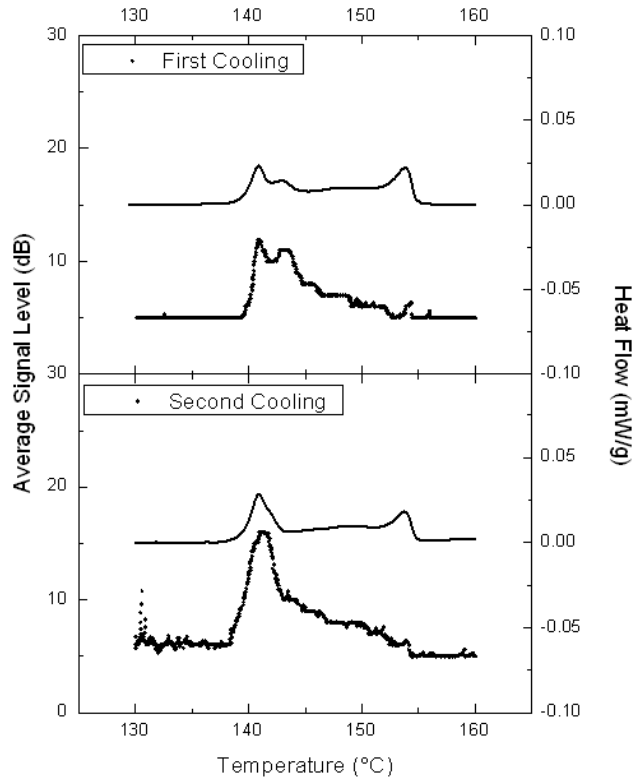

Figure 11. A comparison of DSC (line) and AE (scatter) for the first and second cooling cycles of the water-quenched large-grained specimen, which was compressed 5% and recrystallized at 773 K (500 °C) for one hour.

As a final observation, the appearance of the melting point depression in Figures 8 and 9 may be evidence of the formation of metastable cubic indium, which would be expected to have a lower melting temperature. Cubic indium has been reported in melt-spun Al-In alloys by Van Aken and Fraser [19] and others [20,21]. DSC of melt-spun Al-12In ribbons is shown in Figure 12 and shows three distinct melting distributions: metastable cubic indium, equilibrium melting of tetragonal indium, and the elevated temperature melting of faceted cubic indium particles less than 30 nm in diameter. The crystalline nature of the indium is not expected to play an important role in AE or matrix relaxation as the indium melts and is mentioned here only to provide a better understanding of the melting temperature distribution shown in Figures 8 and 9.

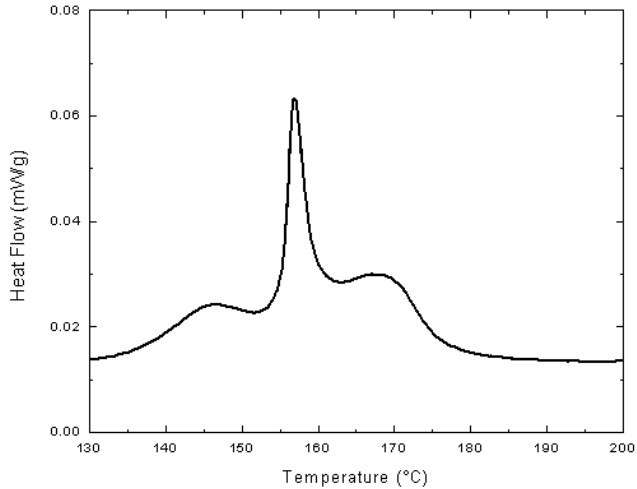

Figure 12. DSC of a melt-spun Al-12In alloy.

## IV. DISCUSSION

*A. General observations of AE and the role of matrix relaxation rate*

Sufficient experimental work has now been presented to support a dislocation generation mechanism for the volume strain relaxation during melting of indium particles embedded in aluminum. The melting temperature of the indium particles is clearly divided into two populations: indium particles embedded within grain interiors and indium particles associated with grain boundaries. Figure 13 shows a summary of all the acoustic energy measurements versus the fraction of indium particles embedded at grain boundaries. Clearly, there is significant scatter in the data, but this should not be surprising since the AE is related to thermal history and related microstructure of the aluminum matrix. In previous studies by Malhotra and Van Aken [12,22,23] both DSC and internal friction measurements were sensitive to prior thermal history. Elevations in melting temperature were observed in both the DSC results and internal friction measurements associated with melting. It was proposed that diffusional relaxation mechanisms accommodated the volume strain upon melting and that long relaxation times would produce melting temperature elevation. Consequently, particles associated with a short matrix relaxation time (on the order of $10^{-6}$ seconds) melt at the equilibrium temperature, and particles associated with a long matrix relaxation time (on the order of $10^5$ seconds) melt at elevated temperature [1]. Relaxation times were calculated based on diffusion of vacancies in aluminum and elastic parameters for the indium particle and aluminum matrix [24]. The distribution of melting temperatures is considered a distribution of matrix relaxation times [12].

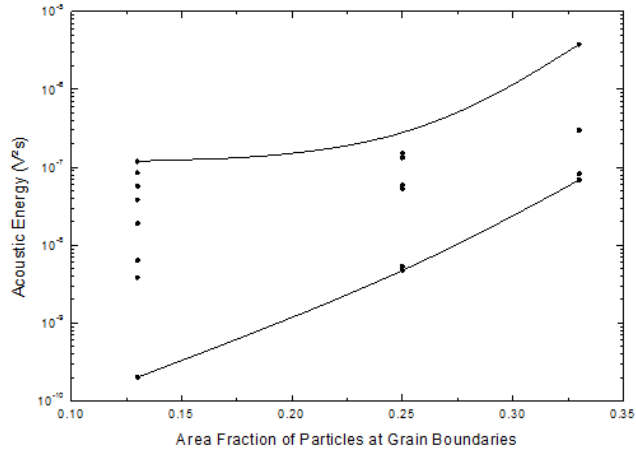

Figure 13. A summary of the acoustic energy as a function of the fraction of particles at grain boundaries.

Kuba and Van Aken have subsequently suggested that prismatic loop generation may be responsible for the equilibrium melting events and this rapid relaxation mechanism would produce AE during particle melting [1]. It remains to be shown why this same dislocation generation mechanism would not operate for indium embedded within the aluminum grains. Dislocation-based strain accommodation would also be reversible and thus compatible with previous internal friction results. That is, the alternating stress during the internal friction measurement affects the stability of the critical nucleus around the melting temperature, with compression stabilizing the solid phase and tension stabilizing the liquid phase.

It is first important to note that microstructural instabilities are expected for aluminum near the melting temperature of indium. As a result, small changes in the dislocation density and substructure near the indium inclusions can affect the rate of matrix relaxation and thus the generation of AE. If prismatic loop punching occurs these defects are not expected to be stable. Frank and prismatic loops in pure aluminum are reported to anneal out between 373 and 473 K (100 and 200 °C), with elimination time as a function of temperature and loop size [25]. Prismatic loops in aluminum are fully annealed out of the microstructure after 10 minutes at 473 K (200 °C) [26], and are very nearly annealed out after 10 minutes at 466 K (193 °C) [27]. Similar studies are reported by Loretto et al. where Frank loops are removed at 411 K (138 °C) after 20 minutes [15]. As an example of the effect of temperature, Edington and Smallman show Frank loops of nearly 0.5 µm in diameter disappear after about 4 minutes at 438 K (165 °C) [28]. Thus, holding the AE specimens at 200 °C should remove any prismatic loop generated by the initial melting transformation.

Evidence for volume strain accommodation by dislocation generation is provided by the 413 K (140 °C) annealing experiments where complete dislocation recovery would be expected prior to the melting experiment. Upon heating the melting event would require generation of new dislocations to accommodate the volume strain of melting and thus produce a stronger AE signal, which was demonstrated in Figure 7 by comparing the annealed AE with the as-cast and the cold worked specimen. The removal of prior dislocation structures would also result in a longer slip distance for the newly-generated dislocations, and thus a stronger AE signal. In this case, the DSC does not change since the fraction of particles melting at the equilibrium temperature does not change. The increase in the number of particles melting at elevated melting temperatures after cold working and low-temperature annealing suggests that the process resulted in shifting a number of particles to a slower relaxation mechanism. The decrease in AE confirms this. Furthermore, it has been shown that grain rotation due to yielding in compression deformation would increase the fraction of low-angle grain boundaries that contain indium inclusions, which are expected to exhibit longer relaxation times [12]. The resulting dislocation network from cold-work may also inhibit loop formation, decrease slip distance, and promote faster diffusional relaxation mechanisms which would be competitive with the dislocation generation mechanism. Changes in the AE and DSC response with sequential heating cycles may now be understood. After the initial heating cycle the specimen reached 200 °C, which should be sufficient to significantly anneal out any prismatic or dislocation loops as the indium particles melt. Upon cooling, fresh dislocations would be generated as the particles contract and these dislocation loops may not completely anneal out. Upon reheating the additional dislocation density would contribute to a faster diffusional relaxation rate and lower both the degree of superheat and AE observed. The results shown in Figures 8 and 9 agree with previous DSC results which found less melting temperature elevation after the initial cycle and that the decrease in particle superheat did not change significantly with subsequent heating cycles after the second cycle [12]. Malhotra and Van Aken have shown that a small amount of cold work after repeated cycling produces a DSC distribution similar to the first scan [12]. An explanation is offered based upon the results of Vandervoort; he shows 5% cold reduction will sweep out all dislocation loops in the material, decreasing the dislocation density and producing a structure similar to the original annealed condition [26]. Therefore, the presence of dislocation loops produced from the last cycle upon cooling may account for the differences in AE and DSC on the second heating cycle. It follows that the cycle of dislocation generation on phase change and dislocation annealing during testing saturates after one test cycle.

*B. Cavitation of indium particles upon quenching from elevated temperature*

As previously noted, cavitation has been cited to explain AE observed during the crystallization of polymers [9] and cavitation could be argued to explain the observed AE across the whole solidification range as shown in Figure 11. It should be emphasized, however, that superheating would not be possible in the presence of a void, since a liquid nucleus would perfectly wet the solid-vapor interface. This is clearly not the case as shown in Figures 8 and 9 where a melting point elevation is clearly demonstrated during the second heating cycle. Malhotra and Van Aken have shown that a significant portion of particles still exhibit superheat even after six heating and cooling cycles and that only the first cycle significantly changes the degree of superheating [12]. Evidence for void formation is observed for the strain annealed specimens quenched from 773 K (500 °C) into water. DSC of these specimens show only melting near the equilibrium temperature and AE is clearly absent. Only after sufficient annealing during the first heating cycle is the AE recovered for melting at the grain boundaries. Unfortunately, verification of voids formed by quenching cannot be made by conventional metallographic preparation, as indium corrodes easily and is often added to aluminum to prevent passivation of sacrificial anodes [29]. Consequently, the indium particles corrode during polishing and in air after polishing. Particle pull-out may also be an issue stemming from the low pressures required to adequately polish the Al-In specimens.

Cavitation was further tested by examining the thermodynamic restrictions on void formation during cooling. Following the analysis by Bourgeois et al. [30], the free energy barrier to void nucleation can be estimated by Equation 1. The surface energy, σ, is calculated as $\sigma = \sigma_v - \sigma_{\alpha\beta}/4$, where $\sigma_v = (\sigma_{v\alpha} + \sigma_{v\beta})/2$. The subscript v denotes a surface energy in vacuum, while α denotes the aluminum matrix and ß denotes the indium particle. Values for the surface energy calculations were taken from de Boer [31]. The thermodynamic driving force for nucleation of a void, $\Delta G_V$, is estimated as $k_B T \ln C_V/C_V^{eq}$, where $k_B$ is Boltzmann's constant, $T$ is absolute temperature of solidification, and $C_V/C_V^{eq}$ is the ratio of actual to equilibrium vacancy concentration. Using Bourgeois's estimate of $10^6$ for the vacancy supersaturation ratio results in a lower limit for the activation energy. A lower vacancy concentration would be expected for the slower cooling rates in air as used in this study (0.4 K/s as compared to 100-1000 K/s). The volumetric strain energy of misfit accommodated by the void, $\Delta G_S$, is estimated to be the fraction volume change upon solidification of the indium particle, 0.025. Therefore, a lower bound for the activation energy of void nucleation, ΔG*, can be estimated as 2.6 eV. Bourgeois comments that an activation energy of 0.027 eV would be expected to yield cavitation voids for 50% of the particles; thus, voids would not be expected to form on air-cooling the Al-In alloys from the casting temperature.

$$\Delta G *= \frac{16\pi\sigma^3}{3(\Delta G_V - \Delta G_S)^2} \tag{1}$$

Formation energies of Frank dislocation loops, perfect prismatic dislocation loops, and voids can be calculated using the analysis of Tan et al. [32]. Figure 14 shows the comparison between activation energies for critical defect nuclei, which act as vacancy sinks upon cooling. Perfect loops are more favorable than voids for low vacancy supersaturations ($C_V/C_V^{eq} < \sim 50$) and Frank loops are more favorable than voids for the entire vacancy supersaturation range examined. Void formation is observed to be competitive with Frank loop generation in Figure 14 for high vacancy supersaturations ($C_V/C_V^{eq} > 10^6$) during solidification of the indium particles. At 773 K (500 °C) the vacancy concentration is calculated to be $1 \times 10^{-4}$, and at 298 K (25 °C) the vacancy concentration is calculated to be $1 \times 10^{-11}$ [33]. Thus, a vacancy supersaturation of $1 \times 10^7$ might be expected for the water quenched specimen. Repeating Bourgeois's analysis in Equation 1 for the higher vacancy supersaturation produces a void activation energy comparable to the value calculated for quenched Al-Sn [30]. Void formation was observed for about 50 % of the particles in Al-Sn after quenching, despite the calculated activation energy being above the theoretical limit [30]. For "slow" cooling rates, vacancy supersaturation would remain low and dislocation loop formation would be more likely to accommodate the volume change on solidification of the indium particles. We can thus conclude that cavitation is not responsible for the AE observed upon cooling for the results shown in Figure 11.

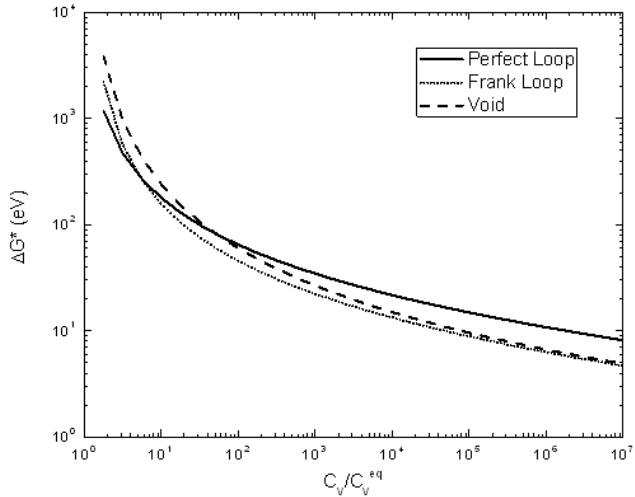

Figure 14. Activation energies for the nucleation of various vacancy sinks as a function of vacancy supersaturation. Calculations were performed at 429 K (156 °C).

If dislocation generation is indeed responsible for volume strain relaxation during melting and solidification it remains to be rationalized why the AE upon solidification is much stronger that that observed upon heating. Particles experiencing undercooling can be considered similar to those experiencing superheat where the matrix behaves rigidly with respect to the volume change of the indium inclusion. Here we must consider the competition between diffusional relaxation mechanisms and the nucleation of new dislocation loops. It has long been recognized that grain boundaries are the principal source for dislocations [34,35]; hence we expect dislocation generation to be relatively easy for particles near grain boundaries and thus produce AE upon melting. It can certainly be argued that not all grain boundaries provide effective sources for dislocation generation and the reader is referred to reference [36].

In contrast, it appears that most embedded liquid indium inclusions generate AE upon cooling with the most undercooled producing the strongest signal. In this case, the molten particle interface may act as a free surface during cooling. Conversely, the particle is solid upon heating and acts as a coupled surface inhibiting dislocation nucleation. Dislocations are easily annihilated at free surfaces; it follows then that the reverse may also be true. Thus, the disproportionately large AE suggests dislocation loop generation at the liquid-solid interface to quickly accommodate the solidifying particle.

*C. A bounding analysis for AE with respect to indium particle size*

Acoustic emission during melting of embedded particles and strain-controlled superheat have only been reported for micron-sized indium particles [1,12,22-24]. AE and matrix relaxation effects might be expected for only a certain size range of particles. Interfacial effects for nanoparticles would control the reaction below this range, and indium particles would melt only at the equilibrium temperature (as though they were bulk indium) above this range. A size range can be established by considering the strain fields around a melting inclusion in an infinite matrix. Bower has derived the matrix strain surrounding an ideal spherical Eshelby particle undergoing a volumetric transformation strain [37]. A 50 nm radius particle melting with 2.5% dilation results in approximately one Burgers vector of elastic displacement in the aluminum matrix. Thus, a lower bound of 100 nm diameter particles is established for strain-dependent transformation and AE.

Elastic accommodation of the critical liquid nucleus within an indium particle can be considered an upper bound. Relaxation via diffusional mechanisms and a lack of AE would be expected, as the lifetime of the critical nucleus would be long. The radius of the critical nucleus is calculated to be 1.5 µm for a superheat of 0.1 K. An indium particle of radius 9 µm would be expected to accommodate that nucleus with approximately one Burgers vector of displacement into the aluminum matrix. Consequently, AE and strain-controlled superheat are precluded for particles outside the range of 0.1 to 18 µm diameter.

It is worth noting that these strain-based calculations were performed using room-temperature data. Near the melting point of indium, the upper bound is expected to decrease due to the easier accommodation of deformation and the lower bound is expected to increase because more strain would be required to nucleate dislocations due to the decrease in Young's modulus. Thus, a narrowing of the predicted range for AE is expected.

*D. Volume strain relaxation as a general phenomenon in solid state phase transformations*

Most solid state phase transformations produce a change in volume and the evidence presented for a dislocation-based accommodation of a volume change without the motion of an interface inspires curiosity into the possible dislocation density produced. Following the analysis of Ahn [38], the number of prismatic loops necessary for an average indium particle to melt can be calculated. For particles larger than approximately 400 times the Burgers vector (about 100 nm for aluminum), the analysis is simply geometric by treating multiple loops as a cylinder of material that is pushed away from the particle. That is, the volume change accommodated by each loop is $\pi r^2 b$, where r is the loop radius (estimated to be 75% of the particle radius) and b is the magnitude of the 1/2<110> Burgers vector for aluminum. The size of the necessary cylinder is calculated as a multiple of the loop's Burgers vector. For the Al-17In alloy previously investigated [1], 73 loops per average particle of 0.33 µm in radius are estimated to be necessary to accommodate the 2.5% volume expansion. If the whole 15.7 wt.% of indium transformed in this manner, the dislocation density would increase by $4.1 \times 10^9$ cm$^{-2}$, which establishes an upper limit. In reality, at most half of the volume of particles typically melt at the equilibrium temperature and display acoustic emission indicative of dislocation formation. Portions of 10 to

20% may be more applicable for some of the specimens presented here depending on the deconvolution of the DSC data. A "typical" microstructure with a theoretical particle size of 1 µm in diameter at the monotectic composition in which 10% of the particles melt at the equilibrium temperature and display AE would generate an increase in dislocation density of 3.5 x $10^8$ cm$^{-2}$. For comparison, continuously cooled bainite is reported to generate a dislocation density of 1.7 x $10^{10}$ cm$^{-2}$ [39]. Van Bohemen [40] has published AE data for continuously cooled bainite. The acoustic emission observed in Al-In and bainite in steels is comparable after normalizing for the volume transformed and the inherent differences in resistance between AE detection systems. The authors conclude that AE cannot be used as a criterion or descriptor of displacive transformations. Rather, any volumetric phase transformation with a short relaxation time may generate AE. In this case, an upper limit to the relaxation time of $10^{-6}$ seconds [1] to 5 x $10^{-5}$ seconds [12] serves as an estimate for AE detection. More sensitive AE systems may detect longer relaxation times.

Kuba and Van Aken [1] had previously concluded after examination of the literature that the aluminum matrix around melting particles on grain boundaries relaxed rapidly and produced AE. They hypothesized the mechanism to be dislocation-based by method of elimination. The current study directly confirms the particles on grain boundaries produce AE and amasses evidence for the dislocation-based relaxation mechanism.

Most solid state phase transformations are noted to be heterogeneously nucleated at grain boundaries. Classical nucleation theory would show that the critical volume of the nucleus is smaller at grain boundaries as a result of surface energy considerations and thus heterogeneous nucleation at grain boundaries has a kinetic advantage. The results of this AE study would suggest that the relaxation of the strain energy at the grain boundary plays as important a role as surface energy. Here the presence of the grain boundary provides the rapid relaxation by nucleating dislocations to accommodate the volume strain of transformation.

## V. CONCLUSIONS

Dislocation emission was determined to be the probable cause of the observed AE in both heating and cooling in Al-In alloys. A size range for the detection of AE and the superheat effect was hypothesized to be 0.1 to 18 µm in diameter. Furthermore, liquid inclusions were suggested to act as a free surface within the higher-melting point matrix and promote dislocation generation during solidification of the indium particles. An upper limit to the dislocation density generated by rapid relaxation and strain accommodation in the aluminum matrix by melting of embedded indium particles was calculated as 4.1 x $10^9$ cm$^{-2}$. Any volume change associated with a diffusion controlled phase transformation may generate AE provided the relaxation of the product or parent phase occurs in less than $10^{-5}$ seconds. Comparisons to continuously cooled bainite suggest that acoustic emission should not be used as a criterion of displacive phase transformations. Strain energy may be as important as surface energy in terms of classical nucleation theory and dislocation generation at grain boundaries may explain the preference in ferrite nucleation along prior austenite grain boundaries in steel.

**ACKNOWLEDGMENTS**

This work was supported in part by the National Science Foundation, the Department of Energy, and the American Iron and Steel Institute under contract No. CMMI 0726888; and Eric Bohannan at the Missouri S&T Materials Research Center is gratefully acknowledged for his help with DSC.

# REFERENCES

[1]     M.M. Kuba and D.C. Van Aken: Mater. Lett., 2012, vol. 77, pp. 89-92.

[2]     G.R. Speich and R.M. Fisher: "Acoustic Emission during Martensite Formation." *Acoustic Emission (ASTM STP 505)*, American Society for Testing and Materials, Philadelphia, 1972, pp. 140-151.

[3]     S.M.C. Van Bohemen, M.J.M Hermans, and G. Den Ouden: Mater. Sci. Technol., 2002, vol. 18, pp. 1524-1528.

[4]     P.C. Clapp: J. Phys. IV, 1995, vol. 5, pp. 11-19.

[5]     B.I. Voronenko: Metal Sci. Heat Treat., 1982, vol. 24, issue 8, pp. 545-553.

[6]     A.L. Purvis, E. Kannatey-Asibu Jr., and R.D. Pehlke: Trans. Am. Foundry Soc., 1996, vol. 103, pp. 1-8.

[7]     H.M. Tensi: Second Acoust. Emiss. Symp. Proc., 1974, pp. 46-57.

[8]     V.B. Vorontsov and V.V. Katalnikov: J. Phys.: Conf. Ser., 2008, vol. 98 (052005), pp. 1-6.

[9]     A. Galeski, L. Koenczoel, E. Piorkowska, and E. Baer: Nature, 1987, vol. 352, pp. 40-41.

[10]    J.R. Frederick, and D.K. Felbeck: "Dislocation Motion As a Source of Acoustic Emission." *Acoustic Emission (ASTM STP 505)*, American Society for Testing and Materials, Philadelphia, 1972, pp. 129-139.

[11]    P.M. Mummery, B. Derby, and C.B. Scruby: Acta Metall. Mater., 1993, vol. 41 (5), pp. 1431-1445.

[12]    A.K. Malhotra, and D.C. Van Aken: Phil. Mag. A, 1995, vol. 71 (5), pp. 949-964.

[13]    A. Wolfenden, and W.H. Robinson: Acta Metall., 1977, vol. 25, pp. 823-826.

[14]    J.L. Murray: J. Phase Equilib., 1983, vol. 4 (3), pp. 271-278.

[15]     M.H. Loretto, L.M. Clarebrough, and P. Humble: Phil. Mag., 1996, vol. 13 (125), pp. 953-961.

[16]     G.M. Vyletel, P.E. Krajewski, D.C. Van Aken, J.W. Jones, and J.E. Allison: Scr. Metall. Mater., 1992, vol. 27 (5), pp. 549-554.

[17]     S.M.C. Van Bohemen, J. Sietsma, M.J.M. Hermans, I.M. Richardson: Acta Mater., 2003, vol. 51 (14), pp. 4183-4196.

[18]     W.M. Haynes and D.R. Lide [ed.]: CRC Handbook of Chemistry and Physics. [Internet Version]. CRC, 2011.

[19]     D.C. Van Aken and H.L. Fraser: Int. J. Rapid Solidif., 1988, vol. 3, pp. 199-222.

[20]     Y. Oshima, T. Nangou, H. Hirayama, K. Takayanagi: Surface Sci., 2001, vol. 476 (1-2), pp. 107-114.

[21]     J. Tao, D.C. Van Aken, J.C. Bilello: MRS Proc., 1993, vol. 307, pp. 299-304.

[22]     A.K. Malhotra and D.C. Van Aken: "An Internal Friction Study of Melting in Aluminum-Indium and Aluminium-Lead Alloys." *M3D: Mechanics and Mechanisms of Material Damping (ASTM STP 1169).* American Society for Testing and Materials, Philadelphia, 1992, pp. 262-281.

[23]     A.K. Malhotra and D.C. Van Aken: Metall. Trans. A, 1993, vol. 24, pp. 1611-1619.

[24]     A.K. Malhotra and D.C. Van Aken: Acta Metall. Mater., 1993, vol. 41 (5), pp. 1337-1346.

[25]     P.S. Dobson, P.J. Goodhew, and R.E. Smallman: Phil. Mag., 1967, vol. 16 (139), pp. 9-22.

[26]     R. Vandervoort and J. Washburn: Phil. Mag., 1960, vol. 5 (49), pp. 24-29.

[27]     F. Kroupa, J. Silcox, and M.J. Whelan: Phil. Mag., 1961, vol. 6 (68), pp. 971-978.

[28]     J.W. Edington and R.E. Smallman: Phil. Mag., 1965, vol. 11 (114), pp. 1109-1123.

[29]     I. Gurrappa: J. Mater. Process. Technol., 2005, vol. I66, pp. 256-267.

[30]     L. Bourgeois, G. Bougaran, J.F. Nie, and B.C. Muddle: Phil. Mag. Lett., 2010, vol. 90, issue 11, pp. 819-829.

[31]     F.R. de Boer, R. Boom, W.C.M. Mattens, A.R. Miedema, and A.K. Niessen: *Cohesion in Metals*, Elsevier Science Publishers B.V., North-Holland, 1988, p. 676.

[32]     T.Y. Tan, P. Plekhanov, and U.M. Gösele: App. Phys. Lett., 1997, vol. 70, p. 1715.

[33]     B. von Guérard, H. Peisl, and R. Zitzmann: App. Phys. A, 1974, vol. 3 (1), pp. 37-43.

[34]     L.E. Murr: Mater. Sci. Eng., 1981, vol. 51 (1), pp. 71-79.

[35]     R.Z. Valiev, V.Y. Gertsman, O.A. Kaibyshev: Phys. Status Solidi A, 1986, vol. 97 (1), pp. 11-56.

[36]     M.D. Sangid, T. Ezaz, H. Sehitoglu, I.M. Robertson: Acta Mater., 2011, vol. 59, pp. 283-296.

[37]     A.F. Bower: *Applied Mechanics of Solids*. CRC Press, 2010, pp. 294-299.

[38]     D.C. Ahn, P. Sofronis, and R. Minich: J. Mech. Phys. Solids, 2006, vol. 54, pp. 735-755.

[39]     H.K.D.H. Bhadeshia: *Bainite in Steels,* IOM Communications, London, 2001, p. 29.

[40]     S.M.C. Van Bohemen, M.J.M. Hermans, G. den Ouden, I.M. Richardson: J. Phys. D: App. Phys., 2002, vol. 35, pp. 1889-1894.